# A Distributed Augmented Reality System for Medical Training and Simulation

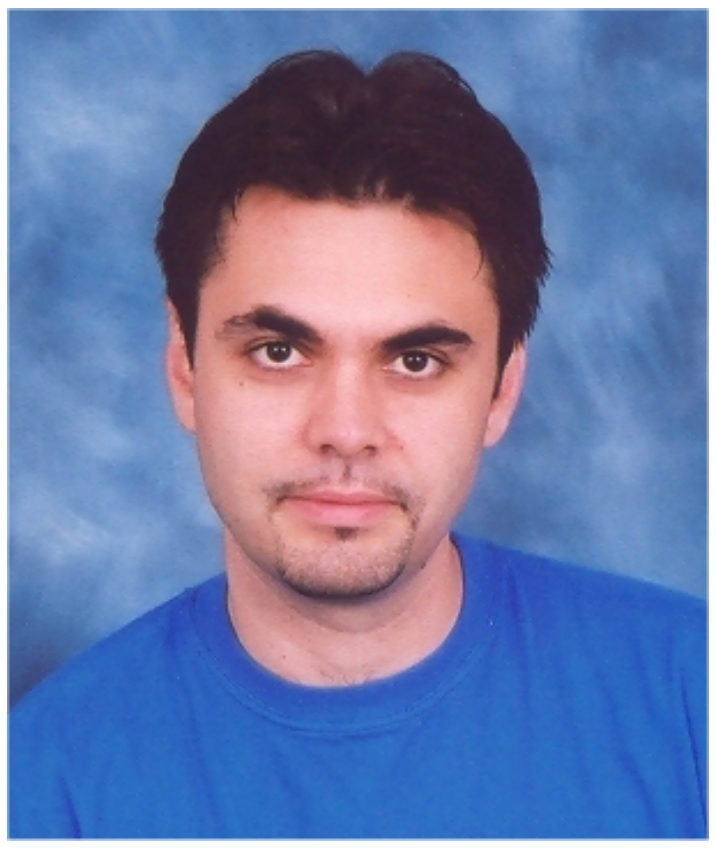

Felix G. Hamza-Lup

School of Computer Science

University of Central Florida

Orlando, Florida 32816-2362

Research Advisors: Jannick P. Rolland and Charles E. Hughes

# ABSTRACT

Augmented Reality (AR) systems describe the class of systems that use computers to overlay virtual information on the real world. AR environments allow the development of promising tools in several application domains. In medical training and simulation the learning potential of AR is significantly amplified by the capability of the system to present 3D medical models in real-time at remote locations. Furthermore the simulation applicability is broadened by the use of real-time deformable medical models.

This work presents a distributed medical training prototype designed to train medical practitioners' hand-eye coordination when performing endotracheal intubations. The system we present accomplishes this task with the help of AR paradigms. An extension of this prototype to medical simulations by employing deformable medical models is possible. The shared state maintenance of the collaborative AR environment is assured through a novel adaptive synchronization algorithm (ASA) that increases the sense of presence among participants and facilitates their interactivity in spite of infrastructure delays.

The system will allow paramedics, pre-hospital personnel, and students to practice their skills without touching a real patient and will provide them with the visual feedback they could not otherwise obtain. Such a distributed AR training tool has the potential to:

- Allow an instructor to simultaneously train local and remotely located students.
- Allow students to actually "see" the internal anatomy and therefore better understand their actions on a human patient simulator (HPS).

## INTRODUCTION

The development of a distributed training tool for endotracheal intubation (ETI) using 3D Augmented Reality (AR) is aimed at medical students, residents, physician assistants, pre-hospital care personnel, nurse-anesthetists, experienced physicians, and any medical personnel needing to perform this common but critical procedure in a safe and rapid sequence. Training a wide range of clinicians in safely securing the airway during cardiopulmonary resuscitation and ensuring immediate ventilation is critical for a number of reasons. First, ETI, which consists of inserting an endotracheal tube (ETT) through the mouth or nose into the trachea, is often a lifesaving procedure. Second, the need for ETI can occur in many places, in and out of the hospital. Perhaps the most important reason for training clinicians in ETI is the inherent difficulty associated with the procedure.

In the case of severe trauma patients, emergency airway management is classified as a major cause of pre-hospital death trauma by the American Heart Association. Moreover, in a 16 hospital study conducted by the National Emergency Airway Registry between August 1997 and October 1998, out of 2392 recorded ETIs, 309 complications were reported, with 132 of these difficulties resulting from intubation procedures [1]. Many anesthesiologists believe that the most common reason for failure of intubation is the inability to visualize the vocal cords. In fact, failed intubation is one of the leading causes of anesthesia-related morbidity and mortality [2]. Thus, there is international concern for the need to effectively train paramedics in pre-hospital emergency situations.

This work presents the development of a distributed medical training prototype designed to train medical practitioners' hand-eye coordination in performing ETI with the help of AR paradigms, and an extension of the prototype to medical simulations that employ deformable medical models. The shared state maintenance of the collaborative AR environment created is assured through a novel adaptive synchronization algorithm which increases the sense of presence among participants and facilitates their interactivity.

The paper is structured as follows. Section two highlights the medical training tools for ETI available today, followed by examples of the application of AR paradigms in the medical field. A brief categorization of the dynamic shared state maintenance techniques for distributed virtual environments is provided. Section three introduces a medical training prototype that uses AR paradigms. The hardware components and the 3D models associated are described, followed by the superimposition procedure and the extension of the simulation with deformable 3D models. Section four highlights the simulation distribution aspects and provides a brief application/infrastructure analysis. Section five introduces the dynamic shared state maintenance algorithm and a preliminary quantitative assessment using quaternion analysis followed by discussions and future work in Section six.

## RELATED WORK

The very first laryngoscope was invented in 1855 by Manuel Garcia, a singing master who wanted to see his own vocal cords. Several years later in 1893, after other inventions, Eisenmenger proposed the cuffed ETT, and the first practical tube produced in 1928 was accepted into common use in the 1930s. These are essentially the tools used today for the ETI procedure. More recently improvements of the ETT have been designed for temporary intubation by unskilled personnel e.g. laryngeal mask [3, 4] or to facilitate intubation e.g. Trachlight [5]. In spite of all these improvements, the need for ETI training is still acute.

Current methods for ETI training include working on cadavers and mannequins. The most widely used model is a plastic or latex mannequin commonly used to teach Advanced Life Support (ACLS) techniques, including airway management. The neck and oropharynx are usually difficult to manipulate without inadvertently "breaking" the model's teeth or "dislocating" the cervical spine, because of the awkward hand motions required. A relatively recent development is the Human Patient Simulator (HPS), a mannequin-based simulator. The HPS is similar to the existing ACLS models, but the neck and airway are often more flexible and lifelike, and can be made to deform and relax to simulate real scenarios. The HPS can simulate heart and lung sounds, and can provide palpable pulses and realistic chest movement. The simulator is interactive, but requires real-time programming and feedback from an instructor [6]. Recently haptic based VR simulators for ETI have been proposed [7] however these developments are still in early stages and suffer from the haptic system instability for long computation cycles.

**Augmented Reality in Medical Applications**

AR systems are used to enhance the perception of the real world. Visually, the real scene a person sees is augmented with computer-generated objects. These virtual objects are placed in the scene (i.e. "registered") in such a way that the computer generated information appears in the correct location with respect to the real objects in the scene. AR can be classified along a virtuality continuum [8] as illustrated in Figure 1.

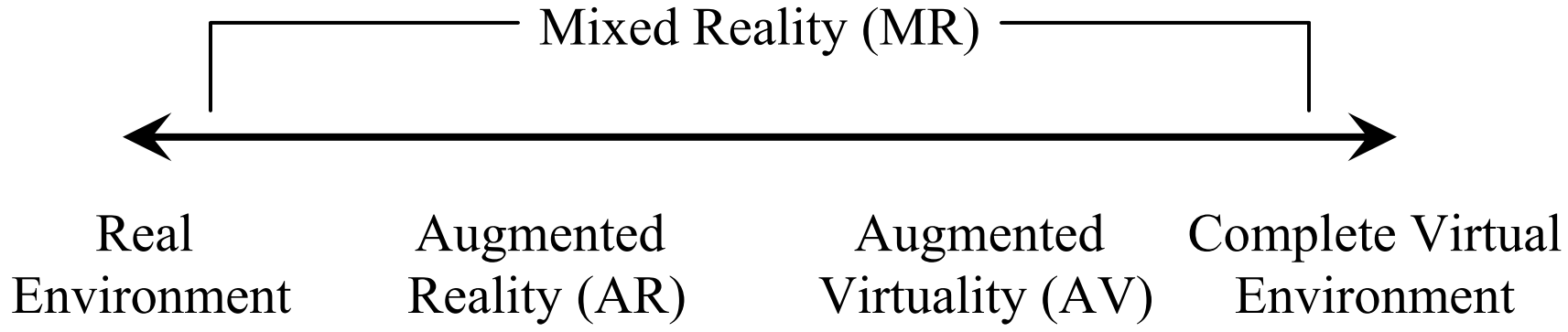

Figure 1. Virtuality Continuum; Real and Virtual worlds can be combined in different proportions.

AR systems were proposed in the mid '90s as tools to assist different fields: medicine [9], complex assembly labeling [10] and construction labeling [11]. One difficulty in AR is the need to maintain accurate registration of the virtual objects with the real world image. This requirement is more stringent in medical applications.

Because imaging technology is so important in the medical field, the AR paradigm was adopted very quickly. Most medical applications deal with image guided surgery [12-14]. Pre-operative imaging studies, such as CT or MRI scans, of the patient provide the surgeon with the necessary view of the internal anatomy. From these images the surgery is planned. Visualization of the path through the anatomy to the affected area is done by first creating a 3D model from the multiple views and slices in a preoperative study. The AR paradigm can be applied such that the surgical team can see the CT or MRI data correctly registered on the patient in the operating theater during the surgical procedure. Other applications of AR in the medical domain are in ultrasound imaging [15] and optical diagnostics [16].

**Distributed AR Based Environments**

With advances in computer graphics, tracking systems, and 3D displays, the research community has shifted attention to distributed environments that use extensively the AR paradigm [17-19].

The main challenge encountered in distributed collaborative AR is the dynamic nature of the environment. The attributes of the virtual components of the scene are changing as an effect of participants' interactions. These interactions and information exchanges generate a state referred to as the "*dynamic shared state*" [20] that has to be maintained consistent at all sites for all participants in the presence of inevitable network latency and jitter. A consistent view in a shared scene may facilitate participant interaction and thus significantly increase the sense of presence among participants [21]. One way in which interaction is related to presence is its ability to strengthen a participant's attention and involvement [22].

A number of consistency maintenance techniques have been employed in distributed virtual environments (VE) and the research efforts can be grouped in four categories. The first category consists of approaches to optimize the communication protocol through packet compression and packet aggregation. The second category includes data visibility management, which makes use of the area of interest management [23] and multicasting [24] and is focused on reducing the bandwidth throughout the system. Taking advantage of the human perceptual limitations, like visual and temporal perception [25] constitutes the third category. The fourth category investigates the system architecture [26].

## A DISTRIBUTED AR MEDICAL TRAINING PROTOTYPE

In an effort to improve airway management training and respiratory system prognostics we have developed the AR based simulator illustrated in Figure 2. Utilizing a HPS from Medical Education Technologies, Inc. (METI) combined with 3D AR visualization of the airway anatomy and the ETT, paramedics will be able to obtain a visual and tactile sense of proper ETI procedure. The system consists of a HPS dressed with retroreflective material, an optical position tracking system, desktop computers, a lightweight head-mounted display (HMD) worn by the participants, and the intubation tools.

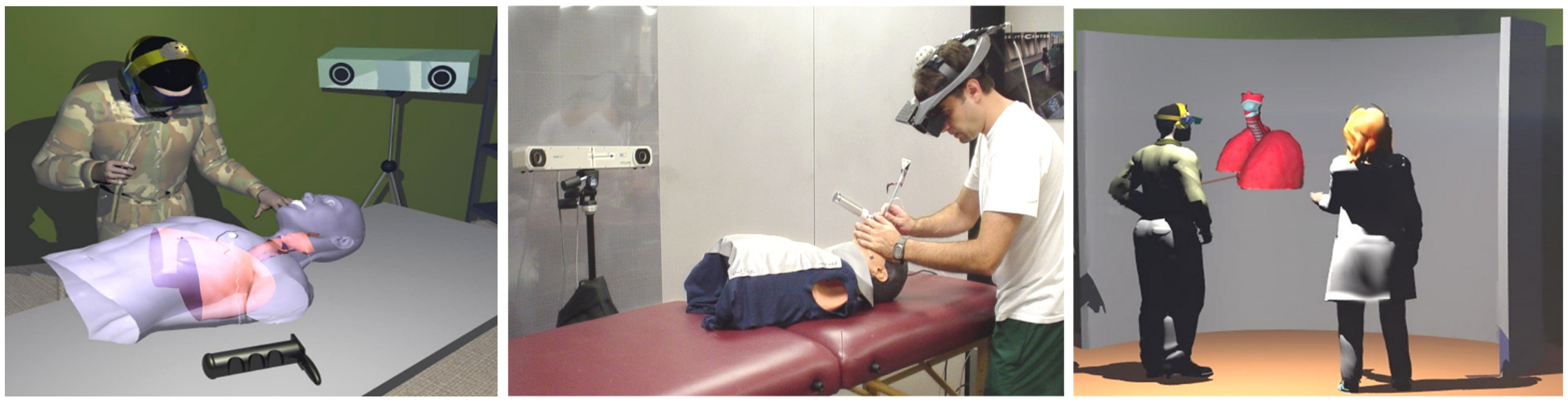

(a) (b) (c)

Figure 2 Illustration of the AR tool for training paramedics on ETI
A trainee (a-concept, b-implementation) performing a medical procedure remotely supervised by the trainer(s) in (c).

In the illustrated scenario (Figure 2.a, 2.b) the trainee sees a virtual 3D model of the internal anatomy superimposed on the HPS. Remotely, an instructor and other trainee(s) wear similar HMDs and are able to visualize the 3D anatomical models while seeing and interacting with each other in a natural way as illustrated in Figure 2.c. As the trainee interacts with the HPS, the relative positions of the 3D models of the ETT, the lungs and the trachea are visualized remotely. The ARC display system shown in Figure 2.c has the potential to allow navigation through the entire virtuality continuum [27].

The system will allow paramedics, pre-hospital personnel and students to practice their skills without touching a real patient and will provide them with the visual feedback they could not otherwise obtain. Such a distributed AR training tool has the potential to:

- Allow an instructor to simultaneously train local and remotely located students.
- Allow students to actually "see" the internal anatomy and therefore better understand their actions on the HPS.

### Hardware System Setup

The AR system integrates a lightweight optical see-through head-mounted display (HMD)[28] illustrated in Figure 3, and an optical tracking system with a Linux-based desktop. A review of the pros and cons of various see-through head-mounted displays for medical applications can be found in [29].

The trainee is able to visualize the internal airway anatomy optically superimposed on the HPS as illustrated in Figure 2.a. With the exception of the HMD, the airway visualization is realized using commercially available hardware components.

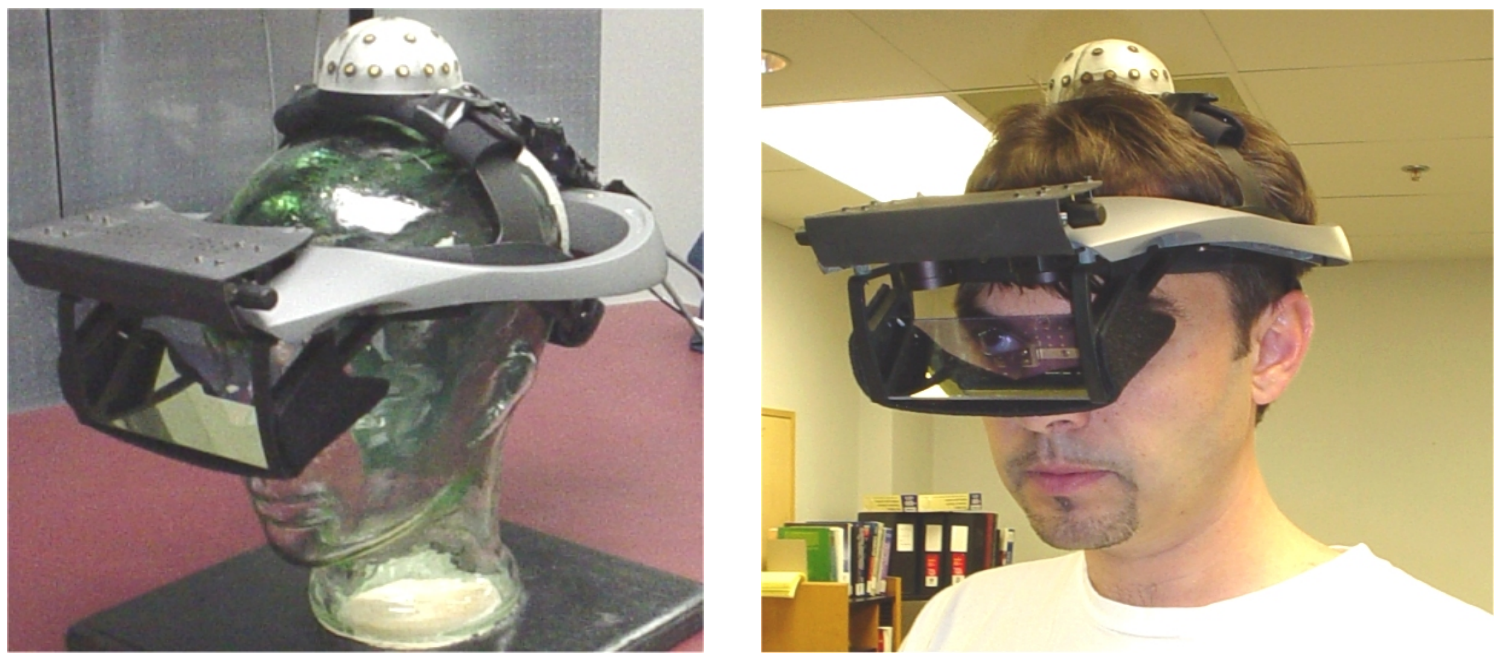

Figure 3. Optical see-through HMD and custom designed semispherical head tracking probe

The pose (i.e. position/orientation) of the trainee's head, the HPS, and the ETT are determined using an optical tracking system (i.e. Polaris™ NDI) and three custom built tracking probes [30, 31] illustrated in Figure 4.

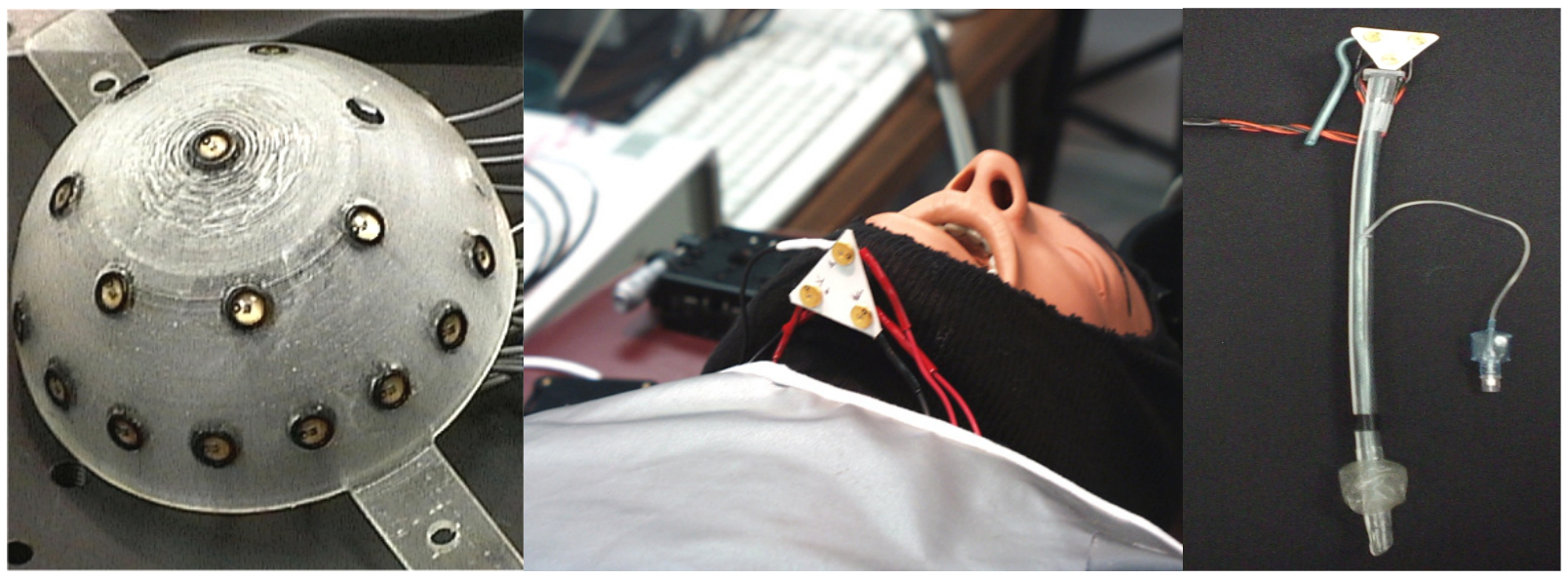

(a) (b) (c)

Figure 4. Collections of LED (i.e. tracking probes): (a) head tracking probe, (b) HPS tracking probe and (c) ETT tracking probe.

The tracking system maximum update rate is 60Hz. Based on a total of three tracking probes, a first one on the HMD to determine the trainee's head position and orientation, a second one on the chin of the HPS to determine its location, and a third one on the intubation tube, the tracking data obtained is currently updated at around 30Hz. The tracking working volume is a cone with 1.5meters height and 0.5meters base radius.

## Anatomically Correct Models for AR: A Challenge

With respect to the airway, the HPS is anatomically correct for a 12-year old boy. If the 3D models are also anatomically correct, the trainee wears the HMD and is able to see the internal anatomy accurately superimposed on the HPS as illustrated in Figure 5. Ideally and ultimately the training tool will provide virtual 3D models of anatomy that exactly correspond to the HPS on which the procedure is performed. The models will also include the capability to deform or add tracheal structures to create difficult intubation events. Such matching between the real and virtual models is critical for the accuracy of registration and for the quality of the simulation.

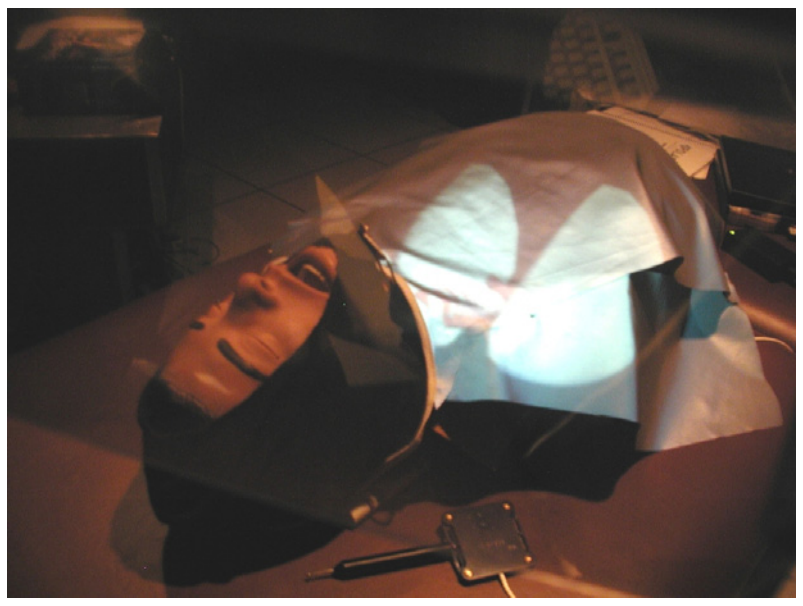

Figure 5. Trainee's view of the static lungs and trachea 3D models superimposed on the HPS

For the prototype implementation, we created in an early implementation a simple 3D anatomical model of the trachea and lungs from 2D models and anatomical statistics. The first model, shown in Figure 6 left, was 665 KB with 1.4 MB of texture.

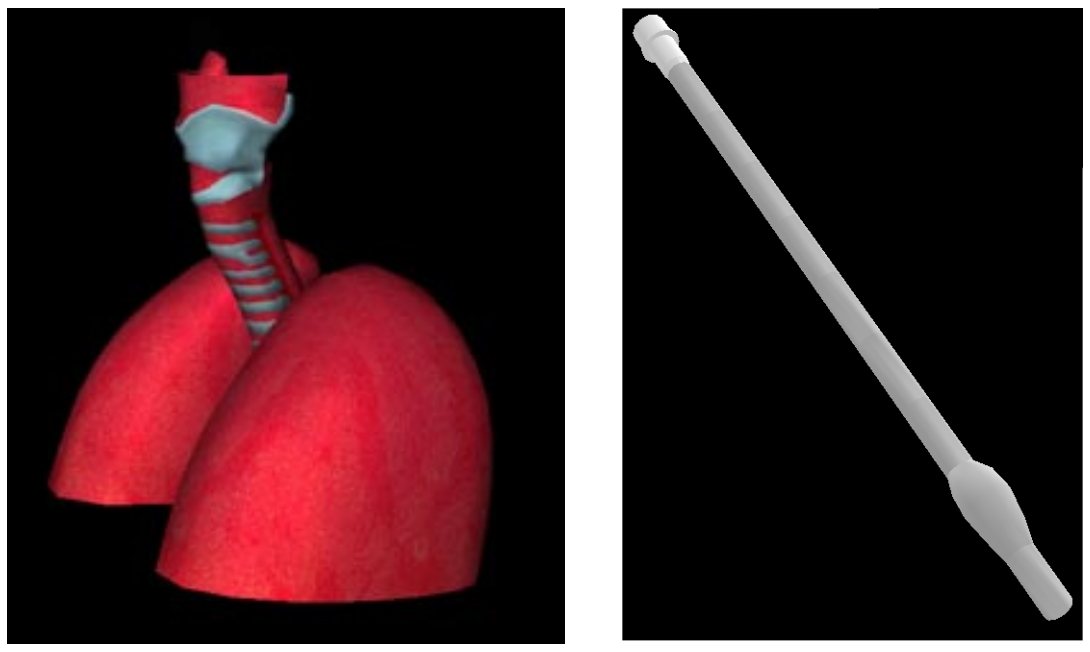

Figure 6. Lungs and Trachea 3D model, ETT (deformable) 3D model

This model was integrated in the first implementation of the prototype given that its relatively small size allows interactive speed rendering. Furthermore, a 3D model of the mandible in correct relation to the trachea proved to be useful in facilitating the registration of the real and virtual models since we can most easily measure the position of the mandible at all times. Once the head is correctly positioned for intubation, we anticipate little motion of the trachea with respect to the mandible. The trachea will then be in correct relationship with respect to the mandible, and the elongation as well as the motion of the trachea during intubation can be modeled.

We are currently working towards integrating 3D anatomical models segmented from the male Visible Human data sets [32]. The first challenge in using these models is their large sizes of 295,276 polygons for the mandible and 1,063,178 polygons for the trachea and mandible together. 3D visualizations of the hand-segmented trachea and mandible from the Visible Human male data sets, generated for this project, are shown in Figure 7. Interestingly, the understanding of the 3D spatial relationships between the anatomical structures, e.g. mandible and trachea, is crucial for planning and understanding the AR based ETI procedure.

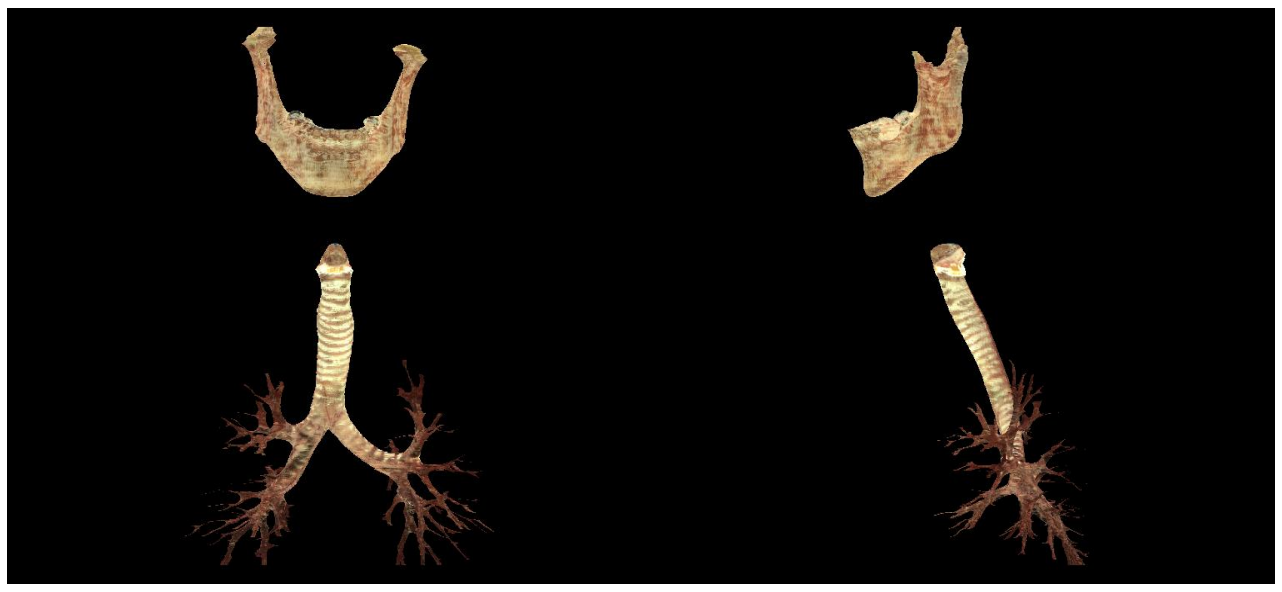

Figure 7. Visualization of bronchial tree. 3D relationship between the trachea and mandible from the Visible Human Male dataset using 3D Vesalius™ Visualizer.

## Superimposition of the Virtual Models on their Real Counterparts

To register (i.e. superimpose) the virtual 3D models of the airway and lungs on the HPS, as well as the 3D model of the ETT on the ETT itself, we have used a least-squares pose estimation algorithm [16].

We compute the transformation from the local coordinate frame of the real object to the virtual object coordinate frame, $\mathbf{M}_{V \leftarrow L}$, using the relationship $\mathbf{y} = \mathbf{Rx} + \mathbf{t}$ where, $\mathbf{y}$ is a virtual object landmark, $\mathbf{x}$ is the corresponding real object landmark, and $\mathbf{R}$, $\mathbf{t}$ represent the rotation and the translation components. Bold letters denote matrices and vectors.

The steps to determine **R** and **t** are:

- Determine the location of virtual models landmarks, $y_i$, where *i* denotes the $i^{th}$ landmark, and the location of the real object (i.e. HPS) landmarks, $x_i$. For the landmarks, we chose four points on the virtual model and the corresponding points on the HPS ribcage. The virtual model landmarks were given directly from the 3D model. To determine the landmark locations on the HPS, we first measured their positions with respect to the tracker using a digitizing probe. From these data, we computed a local coordinate frame for the HPS. The landmark locations were defined with respect to the local coordinate frame.
- Determine the rotation **R** and translation **t** components for the virtual model by minimizing

$$e(R,t) = \sum_{i=1}^{N} \left\| y_i - Rx_i - t \right\|^2 \quad (1)$$

**R** and **t** that minimize the equation best fit a point $x_i$ into $y_i$. **R** and **t** were computed with a method that uses singular value decomposition. The details of the method can be found in [16]

**R** and **t** are then combined to find $\mathbf{M}_{V \leftarrow L}$. It is important to note that the location of the HPS relative to the tracking system does not change during the frame determination and the running of the application. This constraint can be relaxed by attaching a tracking probe on the chin of the patient as illustrated in Figure 4b.

### Simulation Enhancement with Deformable Models

While up to this point the simulator allows superimposition of rigid 3D models, in a real scenario, the patient internal anatomy is non-rigid and deforms based on several factors. Out of a myriad of deformation factors we focused our investigation on the deformation of the lungs.

Physically-based deformation has been applied in areas ranging from animation to surgical simulation. Initial models for such deformations were pioneered by [33, 34]. The integration of a physiological model of the breathing lungs in a distributed simulation is described in [35]. The algorithm for forces distribution and deformation was presented in [36]. The algorithm for lungs deformation allows modeling of various normal and patho-physical conditions [37]. The deformer computes the change in shape based on the volume increase. An illustration of the superimposition of the deformable 3D lungs model is available in Figure 8.

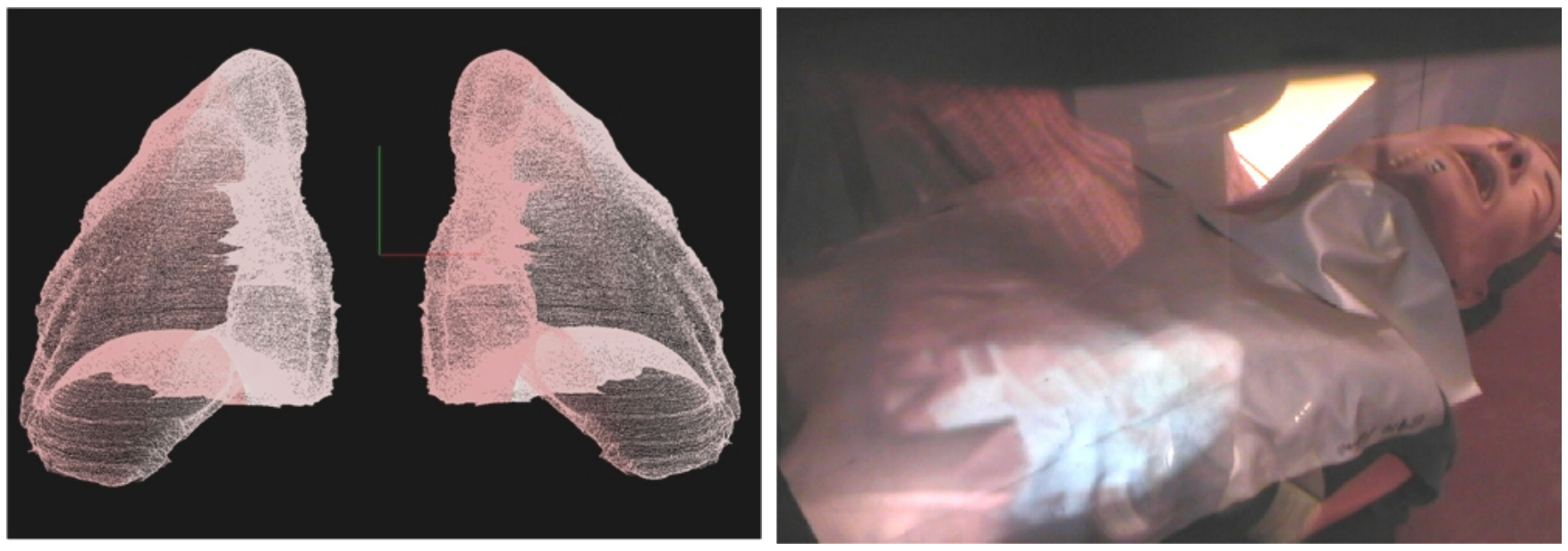

Figure 8. Deformable model of the lungs and the trainee's view of the dynamic (deformable) 3D model of the lungs registered with the HPS

## SIMULATION DISTRIBUTION

An important function of the medical training prototype is its capability to operate remotely. The distributed simulation capability can substantially facilitate experts' interactions, especially during quick-response conditions such as medical emergencies [38]. Distributing the relative position of the 3D models involved in the ETI procedure allows remote visualization. Moreover, augmenting a mannequin based simulator with virtual deformable 3D models that participants may interact with remotely has the potential to enhance training effectiveness allowing trainees to visualize and learn various complex medical procedures with frequent exposure to the techniques and without the cost of travel.

A well designed distributed interactive AR system accounts for the transmission delays of participant actions and provides real-time response. Our recent research focus was the development of shared state consistency maintenance algorithms [39] that will enhance the distributed simulation capability.

Participants make their judgments and react based on the situation presented to them by the human-computer interfaces. In our case, the primary interface is the HMD which allows trainees to see the 3D anatomical models superimposed on the HPS.

In [40] we proposed a novel criterion for categorization of distributed AR based interactive applications by performing an application domain analysis and a deployment infrastructure analysis. In the following sections we will perform this analysis on two distributed applications built using the AR simulator: the distributed ETI training application and the distributed deformable lungs simulation.

### Distributed AR Based ETI Training

The main difficulty in the distributed AR based ETI training application is the data collection and distribution. The data describing the user's head pose (e.g. position and orientation) and the ETT pose is given by the tracking system and is used to register the virtual models with their real counterparts. This data is collected at 30Hz from the optical tracking system and must be distributed immediately such that the remote participants can see these 3D models and their relative position during the ETI procedure.

The corresponding upshot frequency ($\nu_0$), defined in Equation 2 allows us to estimate the capability of the distributed infrastructure where the application will be deployed [40]. In this equation $t_{xy}$ represents the average communication delay between two participants in seconds.

$$\nu_0 = \frac{1\ action}{t_{xy}} \tag{2}$$

We deployed the prototype on a local area network that has an average delay of 1ms. Therefore the infrastructure supports $\nu_0$= 1/0.0001s= 1000 updates/second.

The average action frequency ($\nu_k$) defined in Equation 3 is dependent on the participants' interaction on the virtual models, and *m* represents the number of virtual models affected by the interaction. In the ETI training application (if we consider that the mannequin is static) there are two virtual models: the rigid trachea/lungs and the endotracheal tube (recall Figure 6). $b_{jk}$ represents the number of actions (e.g. updates) applied by participant *k* (i.e. the trainee) on the object *j* (i.e. the virtual 3D models) over a time interval $\Delta t$.

$$\nu_k = \frac{\sum_{j=1}^{m} b_{jk}}{m \cdot \Delta t} \tag{3}$$

In this case the pose of each 3D model is updated at 30 Hz by the tracking system while the trainee manipulates them. In fact, the trachea and lungs 3D model does not change its position. However, in the virtual world its pose will be updated based on the trainer's viewpoint determined by the head tracking probe. The average action frequency, $\nu_k$ will be 30 updates/second. Clearly the distributed interactive AR application falls in the category where the average action frequency is less than the upshot frequency, ($\nu_k < \nu_0$). In this case a high level of shared state consistency can be achieved.

### Distributed Simulation with 3D Deformable Models

In this application scenario each participant's interaction is captured through graphical user interfaces (GUIs). A trainer can change the simulation parameters (i.e. the lungs breathing rate, tidal volume) while remotely the trainee sees the deformable model of the lungs superimposed on the HPS. The trainer's action frequency is limited by the fact that the human-computer response time includes perceptual (i.e., user perceives the items on the display or auditory signals), cognitive (i.e., user retrieves information from his own memory), and motor cycle times which can add up to an average of about 240ms [41]. In other words the average action frequency induced by the participants is $\nu_k$ = 1/0.24 ≈4 (updates/second). The application is deployed over the same network hence $\nu_0$= 1000 updates/second, therefore $\nu_k < \nu_0$.

## DYNAMIC SHARED STATE MAINTENANCE ALGORITHM

The Adaptive Synchronization Algorithm (ASA) proposed is targeted towards distributed interactive applications which fall in the category $\nu < \nu_0$. The algorithm assumes an event-based mechanism, triggered either by the participant actions on the shared scene or by a sensor (e.g. a tracking system) whose update cycle time is comparable or higher than the network latency. Such an assumption is true in the case of our AR based training prototype.

To control the position and orientation of the objects in the shared scene (i.e. seen by all participants) each 3D object has a control packet object (CPO) associated with it. The CPO contains information about the position and orientation of the object, as well as information regarding the actions associated with each object: rotation, translation or scaling. The small size of the CPO ensures low transmission delays. As the CPOs flow through the network, the adaptive synchronization algorithm uses their information to keep the shared scene among different participants consistent. The information carried by the CPOs is distributed to each participating node allowing them to compensate for the network delays.

The synchronization algorithm uses two approaches to *trigger* the information collection. In the first approach, a node measures the round trip time to its neighbor at regular time intervals (e.g. every second). This is the "fixed threshold" approach. Gathering all this data imposes additional overhead at the node and additional network traffic. Moreover these measurements are not required if the network jitter is very low.

An alternative approach consists of *adaptively* triggering the delay measurements for each node, based on the delay history, which better characterizes the network traffic and the interactive application. In the adaptive approach, a fixed threshold is initially used at each node to build the delay history denoted $H_p$. The delay history is a sequence of $p$ delay measurements $h_i$ where $i=1,p$ (the value of $p$ is application dependent; in the implementation we have chosen $p$ to be 100). Furthermore, let $\sigma$ be the standard deviation of $H_p$ and $h_{mean}$ the mean of $H_p$:

$$h_{mean} = \frac{\sum_{i=1}^{p} h_i}{p} \quad \text{and} \quad \sigma = \sqrt{\frac{\sum_{i=1}^{p}(h_i - h_{mean})^2}{p}} \tag{4}$$

Let $h_0$ be the most recent delay, i.e., the last number in the $H_p$ sequence, and $\gamma_0$ the current frequency of delay measurements, expressed as the number of measurements per second. The adaptive strategy is to decrease $\gamma_0$ by 1 unit if $h_0 \in [\, h_{mean} - \sigma,\ h_{mean} + \sigma\,]$ and to increase $\gamma_0$ by 1 if $h_0$ does not belong to this interval. Additional fine tuning is also possible based on the application domain. Details of the algorithm and early results have been reported in [39].

### Algorithm Assessment

To assess the efficiency of the synchronization algorithm, we compute, for a shared 3D object, the amount of orientation drift between the node that acts on the object and each of the other participating nodes. We have focused our experiments on the assessment of the orientation drift. A similar assessment can be done for the object's position.

To emphasize the process we describe a simple scenario. We use two nodes (i.e. a trainee, and a trainer) sharing the same virtual 3D scene. A graphical user interface is available at the trainee, which allows him/her to change the 3D model *orientation* by applying rotations around the Cartesian axes. The trainee generates events from the interface, and each time an event is generated, the object's orientation at both sites is recorded. Because of the network latency, different vectors at each node will describe the orientation of the object. The rotations can be easily expressed using quaternion notation.

Let $q_s$ express the rotation of an object at the trainee's node and let $q_c$ express the rotation of the same object at the trainer's node. Both participants see the same virtual scene and the object should have exactly the same orientation. To quantify the difference between the orientations of the object as rendered on each node, we can compute the correction quaternion $q_E$ every time the user triggers a new action. The correction can be expressed as:

$$q_s = q_E q_c \tag{5}$$

and thus

$$q_E = q_s q_c^{-1} \tag{6}$$

where

$$\alpha = 2\cos^{-1}(\omega_E) \tag{7}$$

The angle α represents the drift between the orientations of the 3D model seen by the two participants. To investigate the effects of the network latency, we performed experiments at different action velocities, given that the drift value for an object is the product of the action velocity applied on that object and the network latency.

For example, let's assume a simple scenario consisting of two nodes representing two participants. Suppose the average delay between these nodes is $t_{ij}$= 0.2 msec and one participant applies an action (e.g. a rotation around axis) on a 3D object in the shared scene with the angular velocity $\omega$ = 10degrees/sec. The angular drift in this case will be given by: $\alpha_{ij} = \omega\, t_{ij} = 0.002$ *degrees*. On a higher delay infrastructure in which the delay is 20 msec, the same action would produce a drift: $\alpha'_{ij} = \omega\, t'_{ij} = 0.02$ *(degrees)*.

While keeping the same delay ($t_{ij}$= 0.2 msec), the drift increase can be simulated by increasing the action velocity, i.e. $\omega$ = 100 degrees/sec. Therefore in order to simulate higher latency networks in our experiments we vary the action angular velocity from 1 degree/sec to 100 degrees/sec. In this way we can simulate network latencies of up to 20 msec on a 0.2 msec average delay local area network.

## Experimental Results

In the first set of experiments we've computed the orientation drift as the participant applies actions on the virtual objects without any compensation. The reason for these measurements is to obtain a reference for the drift magnitude and see its behavior as a participant interacts on the shared scene. The variation of the orientation drift value and its trend line (i.e. third order polynomial fit), while a set of fifty consecutive actions (i.e. random rotations) were applied on the virtual object, is illustrated in Figure 9. Without any compensation the drift accumulates and the drift angle reaches over 210 degrees after 50 consecutive random rotations around the object axes, when the rotations angular velocity is 100 degrees/sec.

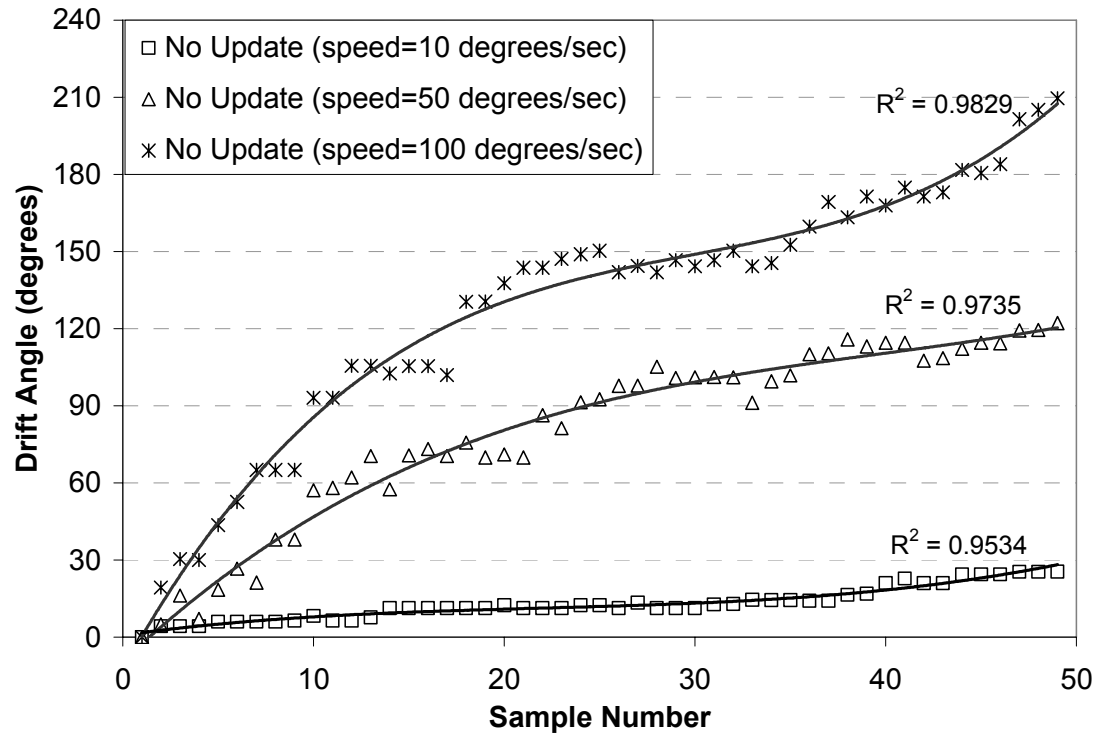

Figure 9. Drift behavior with no drift compensation

In a second set of experiments we update the position of the virtual object after each change in action attributes (e.g. Event Updates). The changes are generated by the participant while interacting on the virtual object from the GUI. As illustrated in Figure 10, the orientation drift is maintained at a fairly constant value. The trend lines use a high order polynomial fit. The drift angle is maintained at an average value of 11 degrees after 50 consecutive random rotations with angular velocity of 100 degrees/sec.

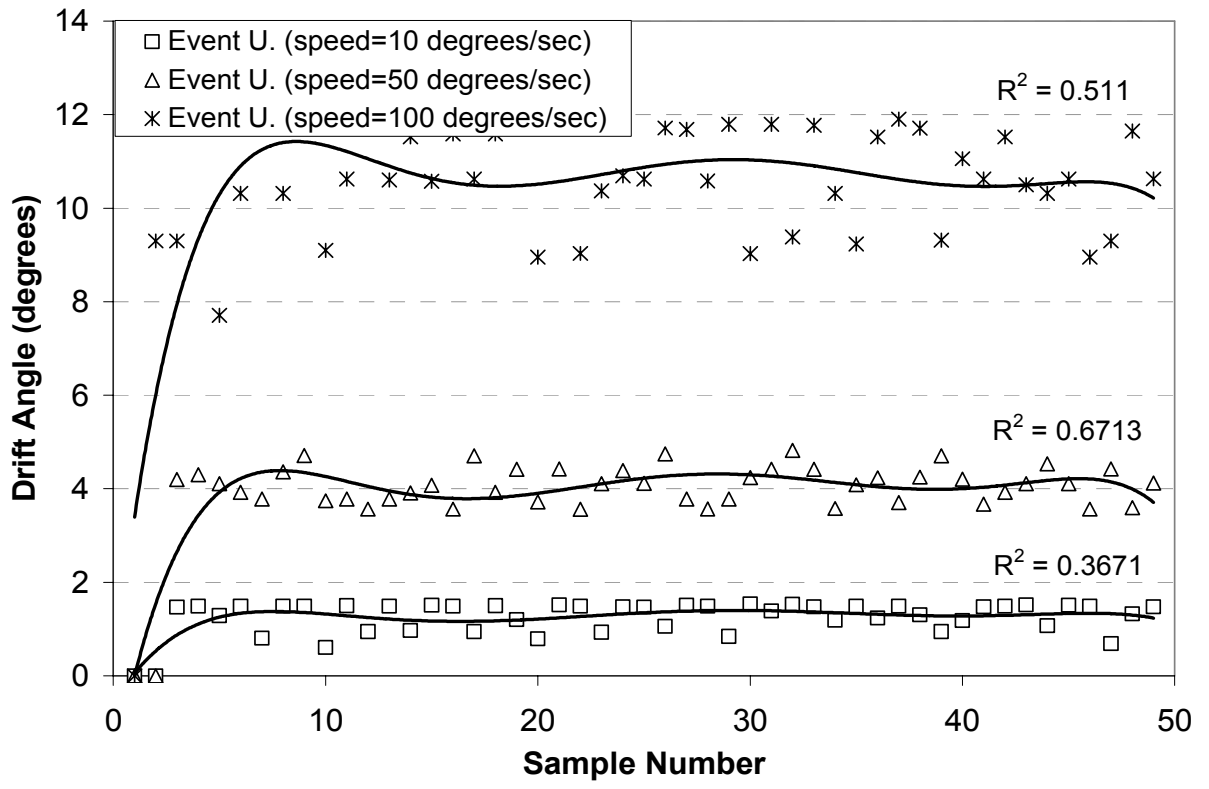

Figure 10. Drift behavior using the Event Updates method

In a third set of experiments, we have employed the adaptive synchronization algorithm to compensate for the communication delay and jitter. As shown in Figure 11, the drift value is significantly decreased and kept at a constant level. A higher order polynomial fit was used for the trend lines. As in the case of event updates, the trend line has a sinusoidal shape which has a negative effect on the correlation coefficient (*R*). The sinusoidal shape of the trend line can be explained as an effect of the buffering and other system threads at the network and operating system level.

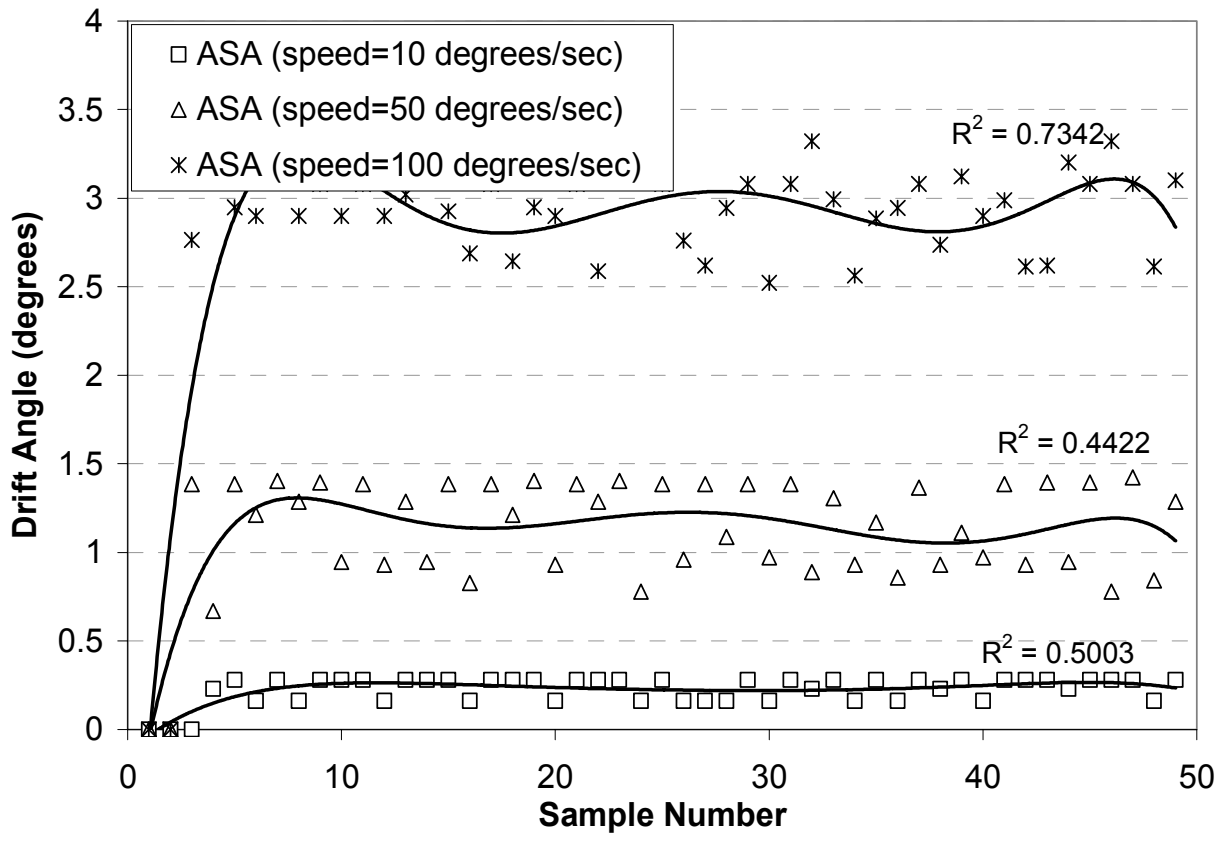

Figure 11. Drift behavior using the Adaptive Synchronization Algorithm (ASA)

When the adaptive synchronization algorithm is used, the drift angle is maintained constant at a value that is two orders of magnitude lower that the average drifts without update and approximately four times smaller that the average drift for the event update approach.

## Scalability: Three, Four, Five and Six-Participant Setup

To start investigating the scalability of the approach, we have increased the number of participants consecutively to three, four, five and six. Figure 12 represents the average orientation drift observed by one participant as the number of participants increases.

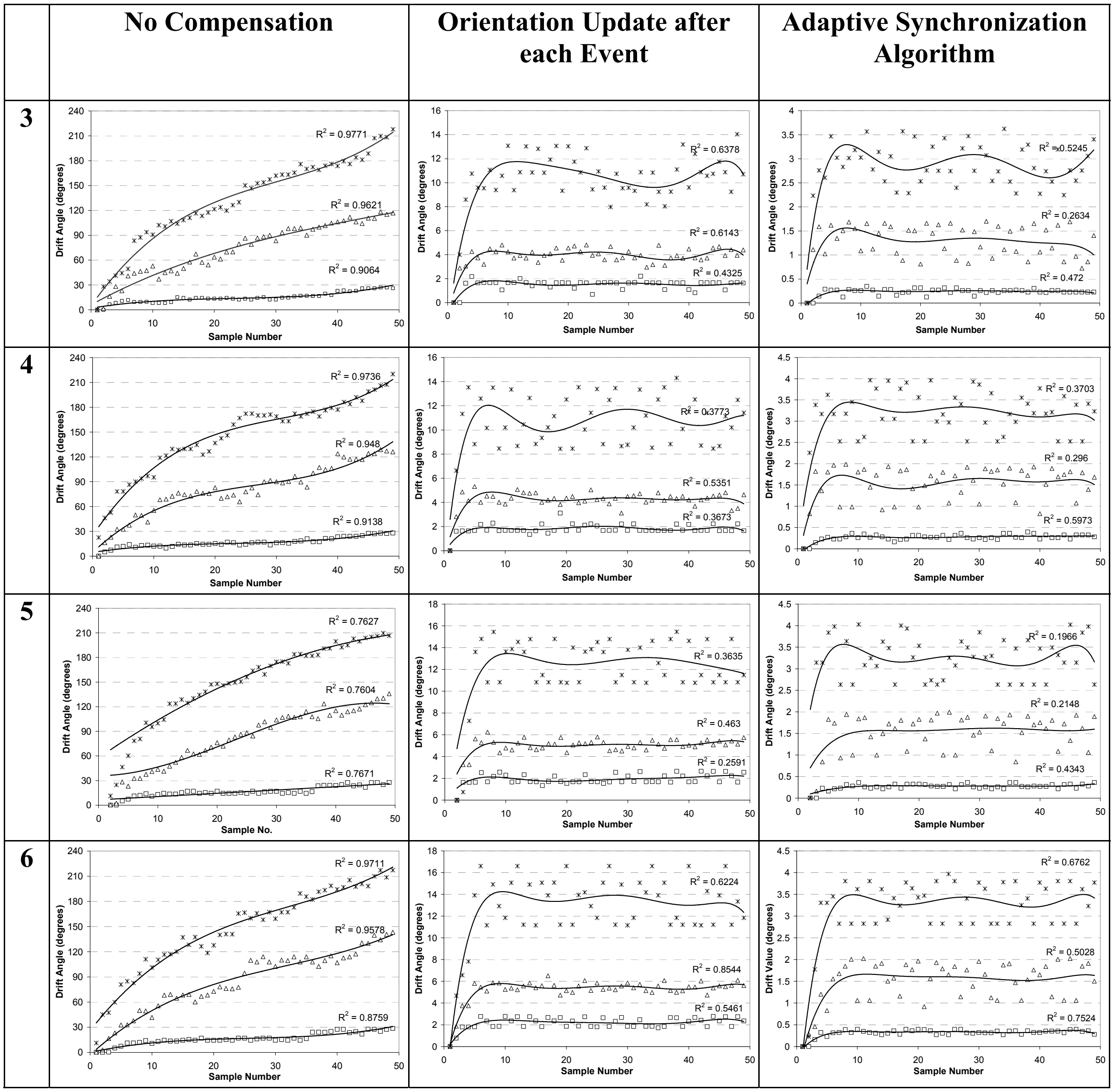

Figure 12. Drift behavior observed by one participant in the 3,4,5 and 6-participant configurations. Leftmost column denotes the number of participants; top row denotes the consistency maintenance strategy.

## Scalability Analysis

To quantify the scalability of the ASA regarding the number of participants, we define a metric analyzing the relationship between the number of participants in the system and the drift values among their views.

Let $\psi_i$ be the average drift value over all the participants, when $i+1$ participants are in the system. Without loss of generality, let us consider an action velocity of 100 degrees/sec. In the case of a two-participant setup, results show that the average drift, $\psi_1$ equals 2.83 degrees, while in the case of a six-participant setup the average drift, $\psi_5$ equals 3.17 degrees. An algorithm with a low degree of scalability would have at least a linear increase in drift, i.e., $\psi_n$ would equal $n^* \psi_1$. On the other extreme, a high degree of scalability would mean $\psi_n \approx \psi_1$. Using this metric, in the six-participant setup, a low degree of scalability would translate to $\psi_5$ equaling $5^*\psi_1$ or 14.15 degrees. However, the experimental results and trend lines in Figure 13 show that $\psi_5 \approx \psi_1$. Thus, the algorithm gives promising results in terms of scalability regarding the number of participants. The trend lines have been plotted using linear regression. The slope of the regression line increases slightly with the action speed.

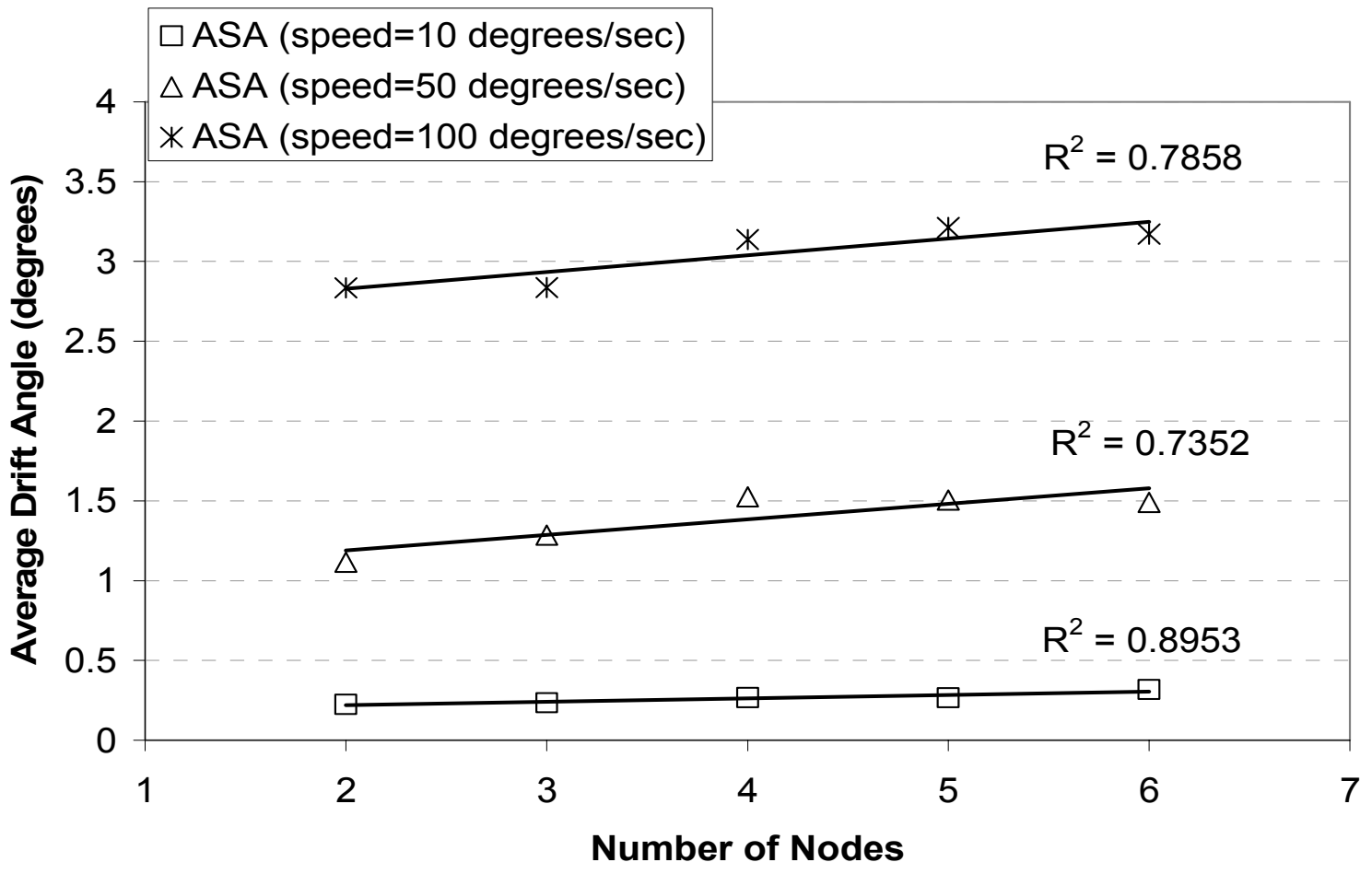

Figure 13. Orientation drift behavior as the number of participants increases for different action speeds

## DISCUSSIONS AND FUTURE WORK

Designing and implementing AR applications for the medical field is a challenging task which requires interdisciplinary research. Additional issues arise in the design and implementation of these applications on a distributed systems infrastructure. The complexity of such systems triggers assessment difficulties. For example, a complex issue is related to the distributed measurements taken for the ASA assessment. These measurements are directly influenced by the accuracy of the NTP [42] synchronization executed before the experiments and point back to the fundamental problem in distributed systems, i.e., the lack of a global clock.

A quantitative investigation of the efficiency of the ASA for the distribution of deformable 3D anatomical models is available in [35]. We would like to stress the fact that there is no general solution for the dynamic shared state maintenance and the approaches taken are application domain dependent. The ASA can be combined with existing shared state management strategies and with convergence algorithms if required.

Finally, the virtual model registration accuracy is another complex issue that needs to be comprehensively investigated. Subjective evaluations of the registration by several participants were conducted, and all participants reported that the rigid as well as the deformable 3D models remained essentially fixed with respect to the HPS chest across multiple viewpoints. A rigorous quantitative study is beyond the scope of this paper.

Future work will include creating different scenarios for training and qualitatively measuring the process for cost-effectiveness, safety, level of physical comfort, and reliability.

## CONCLUSIONS

We have described the overall development and integration of a distributed medical training tool involving AR methods to optically superimpose virtual 3D models on a HPS. The system includes an innovative head-mounted display with light weight optics as well as custom designed motion tracking probes.

We presented an algorithm for dynamic shared state maintenance for a distributed collaborative training environment using such a tool. The adaptive synchronization algorithm (ASA) addresses the impact of network latency on shared scenes in distributed mixed and virtual reality applications. By taking into account the measurement history of the end-to-end network delays among participants, the effect of the network jitter is significantly reduced. The decentralized computation approach for the drift values, carried out independently at each node, improves the system's scalability and its real-time behavior.

## ACKNOWLEDGEMENTS

We wish to thank our sponsors: the Link Foundation, the US Army STRICOM, the NSF/ITR IIS-00-820-16 and the Office of Naval Research for their support for this research. We thank Larry Davis for his contribution to the tracking probes design and implementation as well as Cali Fidopiastis and Anand Santhanam for their contribution to the deformable 3D model of the lungs. We also thank Jay Anton from METI Inc. for stimulating discussions about the project and the HPS. Finally, we thank Celina Imielinska for providing the 3D segmented static model of the lungs from the Visible Human Dataset.